\documentclass[pra,twocolumn,tightenlines,showpacs,nofootinbib]{revtex4-1}
\usepackage{bm,dcolumn,amsmath,graphicx,amsfonts,amssymb}
\usepackage{epsfig}
\newcommand{\bra}[1]{\langle #1|}
\newcommand{\ket}[1]{|#1\rangle}
\newcommand{\threej}[6]{\begin{pmatrix}#1&#2&#3\\#4&#5&#6\end{pmatrix}}
\newcommand{\sixj}[6]{\begin{Bmatrix}#1&#2&#3\\#4&#5&#6\end{Bmatrix}}
\newcommand{\smallspace}{\rule{0pt}{3ex}}
\begin{document} 
\title{Nuclear-spin-dependent parity nonconservation in $s$-$d_{5/2}$ and $s$-$d_{3/2}$ transitions} %
\author{B. M. Roberts}
\author{V. A. Dzuba}
\author{V. V. Flambaum}
\affiliation{School of Physics, University of New South Wales, Sydney,  New South Wales 2052, Australia}
\date{ \today }
\begin{abstract}

We perform calculations of $s$-$d_{5/2}$ nuclear-spin-dependent parity nonconservation amplitudes for  Rb, Cs, Ba$^+$, Yb$^+$, Fr, Ra$^+$ and Ac$^{2+}$.
These systems prove to be good candidates for the use in atomic experiments to extract the so-called anapole moment, a $P$-odd $T$-even nuclear moment important for the study of parity violating nuclear forces.
We also extend our previous works by calculating the missed spin-dependent amplitudes for the $s$-$d_{3/2}$ transitions in the above systems.

\end{abstract}
\maketitle

\section{Introduction}

Parity nonconservation (PNC) in atoms can serve as a very precise low-energy test of the standard model that is a relatively inexpensive alternative to tests performed at high energy (e.g.~at CERN). For more information regarding the history and future prospects of PNC in atoms see, e.g.~\cite{Khr91,GFrev04,DzubaReview2012}.

Currently, the combination of measurements~\cite{meas1,meas2} and calculations~\cite{CsPNC89,CsPNC90,CsPNC92,CsPNC01,CsPNC02,CsPorsev,CsOur}
for the $6s$-$7s$ parity-forbidden E1 transition in cesium provides the most precise atomic PNC result, 
leading to the best atomic test of the electroweak theory so far. 
It is the direct aim of such investigations to determine an experimental value for the nuclear weak charge $Q_W$, a dimensionless coupling constant quantifying the strength of the $Z^0$ exchange between the nucleus and electrons.
The result of this investigation leads to an observed value of the Cs weak charge that
gives a strong indication that improvements and new avenues for investigation in this field could lead to important results~\cite{CsOur,CsQw}.

One way to proceed would be to try to improve the accuracy in both the measurements and calculations in cesium, though it is not expected that significant improvement could be made here in the near future.
Another possibility is to look to other systems. Several proposals have been put forward to search for PNC in heavier atoms, where the PNC signal is expected to be larger (e.g.~\cite{Dzuba2011Yb,Ra,Ba+,KVI,FrPNC}), 
and in systems such as Rb~\cite{OurRb2012}, where the accuracy could be higher.
A promising alternative is to perform measurements of PNC in a chain of isotopes~\cite{IsoChain}, where the accuracy is limited only by the knowledge of the (poorly understood) neutron distribution.

In this work however, we focus our attention on another area, the measurement the $P$-odd $T$-even nuclear moment that arises due to parity violation in the nucleus, the so-called nuclear anapole moment~\cite{AnM,Flambaum1985}.
The experiment~\cite{meas1} of Weiman {\em et al.} 
provides the only measurement of a nuclear anapole moment. 
Measurements of the anapole moment (ANM) could prove to be invaluable tools in the study of parity violation in the hadron sector.

There is interest in measuring PNC in the $6s$-$5d_{5/2}$ transition in cesium~\cite{propose}, and the possibility of measuring PNC in this transition in Ba$^+$ and in the $7s$-$6d_{5/2}$ transition of Ra$^+$ has been discussed~\cite{Geetha1998,Sahoo2011}.
In this work we perform calculations of this and similar amplitudes for several isotopes of 
Rb, Cs, Ba$^+$, Yb$^+$, Fr, Ra$^+$ and Ac$^{2+}$
with the hope of motivating experiment in this important area.
The $s$-$d_{5/2}$ transitions have practically no contribution from the nuclear weak charge, and thus provide good systems for the extraction of the anapole moment.
PNC in $s$-$d$ transitions of moderately charged ions could potentially be measured using techniques put forward by N.~Fortson~\cite{Fortson}.
The prospect of using these elements in measurement of nuclear-spin-independent PNC has been discussed in our recent work~\cite{Fr-like}.

\section{Theory}

The effective Hamiltonian describing the parity violating electron-nucleus interaction 
can be expressed as the sum of the nuclear-spin-independent (SI) and nuclear-spin-dependent (SD) parts 
(unless otherwise stated we use atomic units, $\hbar=|{e}|=m_e=1$, $c=1/\alpha\approx137$ throughout):
\begin{equation}
	 \hat h_{\rm PNC}     =  \hat h_{\rm SI} +  \hat h_{\rm SD} 
	= \frac{G_F}{\sqrt{2}} \left(-\frac{Q_W}{2}\gamma_5 +   \frac{\varkappa}{I}  
 \boldsymbol{\alpha} \boldsymbol{I}\right)\rho(r),
\label{eq:hpnc}
\end{equation}
where $G_F\approx 2.2225\times10^{-14}$ a.u. is the Fermi weak constant, $Q_W$ is the nuclear weak charge, $\boldsymbol{\alpha}=\gamma_0\boldsymbol{\gamma}$ and $\gamma_5= i \gamma_0\gamma_1\gamma_2\gamma_3$ are Dirac matrices, $I$ is the nuclear spin, $\rho(r)$ is the normalised nuclear density, $\int \rho\, {\rm d}^3 r=1$, and $\varkappa$ is a dimensionless constant that quantifies the strength of the SD interaction~\cite{NoteDefinition}.

There are three main sources that contribute to $\varkappa$:
(i) the interaction with the so called anapole moment of the nucleus~\cite{AnM}, this is by far the dominating effect in heavy elements; 
(ii) the contribution from the spin-dependent electron-nucleus weak interaction ($Z^0$ exchange), see e.g.~Ref.~\cite{Flambaum1977};
(iii) the combination of the SI-PNC contribution (i.e.~$Q_W$) with the hyperfine interaction~\cite{Flambaum1985} (see also \cite{Bouchiat1991,Johnson2003}).
The contribution of the combined $Q_W$ and hyperfine effects is discussed in Section~\ref{sec:discuss}.
For greater detail we direct the reader to the review~\cite{GFrev04} and the book~\cite{Khr91}.

The parity-violating ``E1'' transition between two states of the same parity ($a\to b$) is given by the sum
\begin{equation}
\small
E_{\rm PNC} =  \sum_n 
\Big[ 
	\frac{
		\bra{b} \hat{d}_{\rm E1}  \ket{n} \bra{n}  \hat{h}_{\rm PNC}  \ket{a}}
		{E_{a}-E_{n}} 
+	\frac{
		\bra{b}  \hat{h}_{\rm PNC}  \ket{n}  \bra{n}  \hat{d}_{\rm E1}  \ket{a}}
		{E_{b}-E_{n}}
\Big], 
\label{eq:pnc}
\end{equation}
where $\hat{d}_{\rm E1}$ is the electric dipole (E1) operator, 
and $\ket{a}\equiv\ket{J_aF_aM_a}$ with $F=I+J$ the total atomic angular momentum.
With use of the Wigner-Eckart theorem the amplitude can be expressed via the reduced matrix elements:
\begin{equation}
E_{\rm PNC} = (-1)^{F_b - M_b} \threej{F_b}{1}{F_a}{-M_b}{q}{M_a}\bra{J_bF_b}|d_{\rm PNC}| \ket{J_aF_a},
\label{eq:pnc-rme}
\end{equation}
where for the SI amplitude
\begin{align}
&\bra{J_bF_b}| d_{\rm SI}  | \ket{J_aF_a} =
\frac{G_F}{2\sqrt{2}}(-Q_W)(-1)^{I+F_a + J_b+1}  \notag\\
&\times\sqrt{(2F_b+1)(2F_a+1)}\sixj{J_a}{J_b}{1}{F_b}{F_a}{I}  \notag\\
& \times
	\sum_n 
	\Bigg[
		\frac{
			\bra{J_b} |\hat{d}_{\rm E1}|  \ket{J_n} \bra{J_n}  |\gamma_5\rho|  \ket{J_a}}
			{E_{a}-E_{n}} \notag\\
		& \phantom{\times\sum_n	\Big[ \;}
		+
		\frac{
			\bra{J_b} | \gamma_5\rho|  \ket{J_n}  \bra{J_n}  |\hat{d}_{\rm E1}|  \ket{J_a}}
			{E_{b}-E_{n}}
	\Bigg],
\label{eq:si-pnc-rme}
\end{align}
and for the SD amplitude
\begin{align}
&\bra{J_bF_b}| d_{\rm SD}  | \ket{J_aF_a} 	\notag\\
	&=
	\frac{G_F}{\sqrt{2}}\varkappa
  	\sqrt{(I+1)(2I+1)(2F_b+1)(2F_a+1)/I}
	\notag\\
& \times
	\sum_n \Bigg[ (-1)^{J_b-J_a}  
	\sixj{J_n}{J_a}{1}{I}{I}{F_a}
	\sixj{J_n}{J_b}{1}{F_b}{F_a}{I}
	\notag\\
		& \phantom{\times\sum_n	\Bigg[ \;} 
			\times \frac{\bra{J_b} |\hat{d}_{\rm E1}|  \ket{J_n} \bra{J_n}  |\boldsymbol{\alpha}\rho|  \ket{J_a}}
			{E_{a}-E_{n}} \notag\\
		& \phantom{\times\sum_n	\Bigg[ \;} +
(-1)^{F_b-F_a}  
	\sixj{J_n}{J_b}{1}{I}{I}{F_b}
	\sixj{J_n}{J_a}{1}{F_a}{F_b}{I}
		\notag\\
		& \phantom{\times\sum_n	\Bigg[ \;}
			\times\frac{\bra{J_b} | \boldsymbol{\alpha}\rho|  \ket{J_n}  \bra{J_n}  |\hat{d}_{\rm E1}|  \ket{J_a}}
			{E_{b}-E_{n}}\Bigg].
\label{eq:sd-pnc-rme}
\end{align}

In tables we present the $z$-components:
\begin{equation}
E_{\rm PNC}(z) = (-1)^{F_b - F_z} \threej{F_b}{1}{F_a}{-F_z}{0}{F_z}\bra{J_bF_b}|d_{\rm PNC}| \ket{J_aF_a},
\label{eq:pnc-z}
\end{equation}
where we take $F_z={\rm min}(F_a,F_b).$

\section{Calculations} \label{sec:calcs}
If the states $a$, $b$ and $n$ in (\ref{eq:pnc}) are the physical many-electron wavefunctions of the atom then these equations are exact and the summation is over all  excited states.
In calculations obviously this is not the case, we use single-electron orbitals as the wavefunctions and extend the sum over all states (the summation over core states corresponds to including the highly excited autoionization states).

We begin with the relativistic Hartree-Fock (RHF) approximation, generating the single-particle orbitals in a $V^{N-1}$ potential.
Core-valence correlation effects are then included using the correlation potential (CP) method~\cite{CPM}, and the polarization of the core electrons and interactions with external fields are taken into account using the time-dependent Hartree-Fock (TDHF) approximation~\cite{CPM,CPM2}.

The correlation potential, an {\em ab initio}, non-local (integration), energy-dependent operator, $\hat\Sigma=\hat\Sigma(E,l,j)$, is calculated using a summation of dominating diagrams of many-body perturbation-theory (including screening of the electron-electron interaction and the particle-hole interaction) to all orders using the Feynman diagram technique and relativistic Hartree-Fock Green's functions~\cite{CPM2}. 
Then by solving the relativistic Hartree-Fock-like equations with the extra operator $\hat \Sigma$,   
\begin{equation}
 (\hat H_0 +\hat \Sigma - \epsilon^{(\rm BO)}_n)\psi_n^{(\rm BO)}=0,
\label{eq:BO}
\end{equation}
we construct the Brueckner orbitals (BOs) for the valence electron.
Here, $H_0$
 is the RHF Hamiltonian and the index $n$ denotes valence states. 

Part of the missing diagrams can be expressed in terms of the energy derivatives of $\Sigma$, or can also be calculated separately. 
These contributions are very small for alkaline atoms but may be significant in atoms where radius of valence electron is close to the core electron radius (e.g.~in Yb$^+$).
 The correlation potential method is especially accurate in atoms with one electron above closed subshells (which are the topic of the present work), where it gives an accuracy of about $\sim0.1\%$ for the ionization energies of valence electron orbitals. 

Note that the  correlation potential is calculated independently
for  orbitals with different $l$, $j$.  Therefore, we  may estimate the missing (and very small) contributions of higher order diagrams by using a simple semi-empirical procedure of rescaling the CP operator, i.e.~$\hat\Sigma\to\lambda\hat\Sigma$ in Eq.~(\ref{eq:BO}).
A different parameter is chosen for each partial wave (i.e.~$ns$, $np_{1/2}$, $np_{3/2}$, $nd_{3/2}$, and $nd_{5/2}$) to reproduce exactly the experimental energies corresponding to the lowest (valence) principal quantum number for each partial wave.
It should also be noted that these parameters typically differ from 1 by only a small fraction, e.g.~for cesium they are
$\lambda_{s}$=$0.99$, $\lambda_{p_{1/2}}$=0.96, $\lambda_{p_{3/2}}$=0.97, $\lambda_{d_{3/2}}$=0.94, and $\lambda_{d_{5/2}}$=0.94, indicative of the already very good accuracy of the {\em ab initio} all-order CP method.

This fitting makes only a small difference to most PNC amplitudes, and the difference between amplitudes calculated with and without the fitting provides a good indication of the relative size of any missed correlations and thus serves as a good estimate of the uncertainty. In the transitions here however, the uncertainty is dominated by core-polarisation effects, not the correlation potential.
It is also important to note that even in cases where this fitting does make a difference its effect on the ratio of the SI to SD parts is negligible.

For Yb$^+$ we use only the second-order CP method due to the more complicated electron structure.
The presence of the $4f^{14}$ shell means there are other correlation effects that are larger than the all-order corrections, see e.g.~Ref.~\cite{Dzuba2011}.
The second order CP operator provides reasonable accuracy as is, and the process of rescaling means the accuracy is good here also, as discussed in the next section.

In the evaluation of the amplitude, the operators $\hat{d}_{\rm E1}$ and $\hat{h}_{\rm PNC}$ in Eq.~(\ref{eq:pnc}) are modified to include the effect of the polarization of the core electrons due to the interaction with the external E1 and weak fields: 
$\hat{d}_{\rm E1} \rightarrow \hat{d}_{\rm E1} + \delta V_{\rm E1}$ and 
$\hat{h}_{\rm PNC} \rightarrow \hat{h}_{\rm PNC} + \delta V_{\rm PNC}$. 
Here $\delta V_{\rm E1}$ ($\delta V_{\rm PNC}$) is the modification to the RHF potential due to the effect of the external field $\hat{d}_{E1}$ ($\hat h_{\rm PNC}$). 
In the TDHF method, the single-electron orbitals are perturbed in the form $\psi=\psi_0+\delta\psi$ where $\psi_0$ is an eigenstate of the RHF Hamiltonian, and $\delta\psi$ is the correction due to the external field.
The corrections to the potential are then found by solving the set of self-consistent TDHF equations for the core states:
\begin{equation}
(\hat H_0 -\varepsilon_c)\delta \psi_{c} = -(\hat f +\delta V_f)\psi_{0c},
\label{eq:RPA}
\end{equation} 
where the index $c$ denotes core states and $\hat f$ is the operator of external field (be that $\hat{d}_{E1}$ or $\hat h_{\rm PNC}$).

Note that the approach described above does not take into account the effect of core polarization due to simultaneous action of the weak and E1 fields. 
This `double-core-polarization' (DCP) effect was the study of our recent work, Ref.~\cite{DCP}.
Accurate calculations would require the use  of the `solving equations'  approach (see e.g.~\cite{CsPNC02}), a more numerically stable method based on solving differential equations, which includes the DCP contribution. 
However, since high accuracy is not needed for the SD-PNC, we use simpler approach which is based on a direct summation over states.
We use Ref.~\cite{DCP} to include the DCP correction into the SI amplitudes, but do not include this term into the SD amplitudes since the accuracy of analysis is less important here.

To use the direct-summation method, we employ the B-spline technique~\cite{B-spline} to construct the set of single-electron orbitals used for the summation in Eq.~(\ref{eq:pnc}), as well as for the calculation of $\hat \Sigma$.
The states used in the  calculation of $\hat \Sigma$ are linear combinations of the B-splines which are
eigenstates of the RHF Hamiltonian, whereas those used for the evaluation of (\ref{eq:pnc}) are the Brueckner orbitals (eigenstates of the $\hat H_0 + \hat \Sigma$ Hamiltonian).
For the summation we use 90 B-splines of order 9 for each partial wave in a cavity of radius 75~$a_0$.

\subsection{Accuracy of the calculations}

  \begin{table}%
    \centering%
    \caption{Calculated ionization energies for cesium in various approximations and comparison with experiment (Ref.~\cite{NIST}).
Blank means calculated value matches exactly with experiment by construction. Units: cm${}^{-1}$.
} 
\begin{ruledtabular}%
  \begin{tabular}{rrrrrr}
  \multicolumn{1}{c}{Level	}
& \multicolumn{1}{c}{ $\Sigma^{(2)}$	}
& \multicolumn{1}{c}{ $\lambda\Sigma^{(2)}$	}
& \multicolumn{1}{c}{ $\Sigma^{(\infty)}$	}
& \multicolumn{1}{c}{ $\lambda\Sigma^{(\infty)}$	}
&  Exp.   \\
\hline
$6s_{1/2}$	&  -32416	&  	&  -31457	&  	&  -31406   \\
$6p_{1/2}$	&  -20539	&  	&  -20290	&  	&  -20228   \\
$6p_{3/2}$	&  -19940	&  	&  -19722	&  	&  -19674   \\
$5d_{3/2}$	&  -17567	&  	&  -17146	&  	&  -16907   \\
$5d_{5/2}$	&  -17407	&  	&  -17030	&  	&  -16810   \\
$7s_{1/2}$	&  -13024	&  -12832	&  -12827	&  -12817	&  -12871   \\
$7p_{1/2}$	&  -9710	&  -9628	&  -9640	&  -9624	&  -9641   \\
$7p_{3/2}$	&  -9521	&  -9448	&  -9458	&  -9445	&  -9460   \\
$8p_{1/2}$	&  -5724	&  -5689	&  -5694	&  -5687	&  -5698   \\
$8p_{3/2}$	&  -5639	&  -5607	&  -5611	&  -5606	&  -5615   \\

 \end{tabular}%
\end{ruledtabular}%
    \label{tab:EnvsExp}%
  \end{table}%


Without any rescaling of the correlation potential (see Section~\ref{sec:calcs}) 
our energies agree with experiment to around 0.1\%-0.5\% for most levels, and the important $s$-$p$ intervals are reproduced to about 0.3\%. 
A detailed analysis of the accuracy in these systems has also been performed in our recent papers, Ref.~\cite{Fr-like,Dzuba2011}, where we present calculations for the same atoms and ions investigated here.
In Table~\ref{tab:EnvsExp} we present calculated energy levels for cesium using the second-order  ($\Sigma^{(2)}$) and the all-order ($\Sigma^{(\infty)}$) CP method, both with and without scaling.
Table~\ref{tab:intervalvsExp} presents the percentage discrepancies for the relevant energy intervals in cesium. 
This shows the small effect that scaling has directly on the energies, but the relatively large improvements it makes on the intervals.
The rescaling of the correlation potential helps to numerically stabilize the results.
The rescaling means there is no significant loss in the accuracy for the energy-levels when using $\Sigma^{(2)}$ instead of $\Sigma^{(\infty)}$.
This is important for the case of Yb$^+$, where only the second-order correlation potential was used.

  \begin{table}%
    \centering%
    \caption{Percentage variation between the experimental (from~\cite{NIST}) energy intervals of relevance to parity nonconservation in cesium and calculations in various approximations.
Blank means calculated value matches exactly with experiment by construction.} 
\begin{ruledtabular}%
  \begin{tabular}{ldddd}
\multicolumn{1}{c}{Interval}
&  \multicolumn{1}{r}{$\Sigma^{(2)}$}	
&  \multicolumn{1}{r}{$\lambda\Sigma^{(2)}$}	
&  \multicolumn{1}{r}{$\Sigma^{(\infty)}$}	
&  \multicolumn{1}{r}{$\lambda\Sigma^{(\infty)}$}	   \\
\hline
$6s_{1/2}-6p_{1/2}$	&  6.35	&  	&  -0.10	&     \\
$6s_{1/2}-6p_{3/2}$	&  6.31	&  	&  0.02	&     \\
$6s_{1/2}-7p_{1/2}$	&  4.08	&  -0.02	&  0.24	&  0.08   \\
$6s_{1/2}-7p_{3/2}$	&  4.08	&  -0.01	&  0.24	&  0.07   \\
$5d_{3/2}-6p_{1/2}$	&  -5.17	&  	&  -5.35	&     \\
$5d_{3/2}-6p_{3/2}$	&  -7.32	&  	&  -6.92	&     \\
$5d_{3/2}-7p_{1/2}$	&  4.83	&  -0.05	&  3.30	&  0.24   \\
$5d_{3/2}-7p_{3/2}$	&  4.81	&  -0.04	&  3.24	&  0.20   \\
$5d_{5/2}-6p_{3/2}$	&  -5.54	&  	&  -6.02	&     \\
$5d_{5/2}-7p_{3/2}$	&  4.28	&  -0.04	&  3.02	&  0.20   \\
 \end{tabular}%
\end{ruledtabular}%
    \label{tab:intervalvsExp}%
  \end{table}%

  \begin{table*}%
    \centering%
    \caption{Calculations of reduced matrix elements (a.u.) of electric dipole transitions of interest to PNC studies in cesium and comparison with experiment. The last column shows the percentage difference between final calculations (using the rescaled all-order correlation potential, $\lambda\Sigma^{(\infty)}$) and experiment. } 
\begin{ruledtabular}%
  \begin{tabular}{ldddddld}
& \multicolumn{4}{c}{This work } 
& \multicolumn{3}{c}{ Experiment}  \\
\cline{2-5}\cline{6-8}
  \multicolumn{1}{c}{Transition}
& \multicolumn{1}{r}{ $\Sigma^{(2)}$	}
& \multicolumn{1}{r}{ $\lambda\Sigma^{(2)}$	}
& \multicolumn{1}{r}{ $\Sigma^{(\infty)}$	}
& \multicolumn{1}{r}{ $\lambda\Sigma^{(\infty)}$	}
& \multicolumn{1}{c}{ Value } 
& \multicolumn{1}{c}{ Ref. } 
& \multicolumn{1}{c}{ \% Diff. }  \\
\hline
$6s_{1/2}-6p_{1/2}$	&  4.387	&  4.503	&  4.506	&  4.512	&  4.4890(65)	&  \cite{rafac}	&  0.51 \\
	&  	&  	&  	&  	&  4.5097(74)	&  \cite{young}	&  0.05 \\
$6s_{1/2}-6p_{3/2}$	&  6.170	&  6.337	&  6.343	&  6.351	&  6.3238(73)	&  \cite{rafac}&  0.42 \\
	&  	&  	&  	&  	&  6.3403(64)	&  \cite{young}		&  0.16 \\
$6s_{1/2}-7p_{1/2}$	&  0.2995	&  0.2744	&  0.2645	&  0.2724	&  0.2757(20)	&  \cite{vasilyev}	&  1.19 \\
&  	&  	&  	&  &  0.2825(20)	&  \cite{shabanova,vasilyev}&  3.56 \\
		$6s_{1/2}-7p_{3/2}$	&  0.6050	&  0.5686	&  0.5581	&  0.5659	&  0.5795(100)	&  \cite{vasilyev}	&  2.34 \\
	&  	&  	&  	&  	&  0.5856(50)	&  \cite{vasilyev}	&  3.36 \\
$5d_{3/2}-6p_{1/2}$	&  6.744	&  7.039	&  6.927	&  7.032	&  7.33(6)	&  \cite{diBerardino}	&  4.07 \\
$5d_{3/2}-6p_{3/2}$	&  3.037	&  3.173	&  3.121	&  3.170	&  3.28(3)	&  \cite{diBerardino}	&  3.37 \\
$5d_{5/2}-6p_{3/2}$	&  9.254	&  9.629	&  9.481	&  9.616	&  9.91(3)	&  \cite{diBerardino}	&  2.97 \\
 \end{tabular}%
\end{ruledtabular}%
    \label{tab:MEvsExpCs}%
  \end{table*}%


In Table~\ref{tab:MEvsExpCs} we compare calculations of several of the relevant $E1$ reduced matrix elements  for cesium with their corresponding experimental values.
This shows very good agreement with experiment, to better than 0.5\% for the lowest $s$-$p$ transitions, and better than 5\% for the transitions involving $d$ and higher $p$ states.
Again, we present calculations using the second-order ($\Sigma^{(2)}$) and the all-order ($\Sigma^{(\infty)}$) CP method, both with and without scaling.
We demonstrate that by including the rescaling of the correlation potential we can correct for the discrepancies that arise from using the second-order correlation potential, effectively meaning that the rescaled second order CP is practically as good as using the all-order method.

  \begin{table}%
    \centering%
    \caption{Calculated reduced matrix elements (a.u.) for electric dipole transitions of interest in Ba$^+$ and Yb$^+$ and comparison with experiment where available.} 
\begin{ruledtabular}%
  \begin{tabular}{lllllll}
& \multicolumn{3}{c}{Ba$^+$} 
& \multicolumn{2}{c}{Yb$^+$}  \\
\cline{2-4}\cline{5-7}
  \multicolumn{1}{c}{Transition}
& \multicolumn{1}{c}{Calc.}
& \multicolumn{2}{c}{ Exp.	}
& \multicolumn{1}{c}{Calc.}
& \multicolumn{2}{c}{ Exp.	} \\
\hline
\smallspace
$6s_{1/2}-6p_{1/2}$	&  3.322	&  3.36(4)	&  \cite{Davidson}	&  2.705	&  2.471(3)	&  \cite{Olmschenk} \\
$6s_{1/2}-6p_{3/2}$	&  4.690	&  4.55(10)	&   \cite{Davidson}	&  3.817	&  3.36(2)	&  \cite{Pinnington} \\
$5d_{3/2}-6p_{1/2}$	&  3.063	&  3.03(9)	&  \cite{Kastberg}	&  3.094	&  2.97(4)	& \cite{Olmschenk} \\
								&  	&  3.14(8)	&  \cite{Sherman}	&  	&  	&   \\
	&  	&  2.90(9)	&   \cite{Davidson}	&  	&  	&   \\
$5d_{3/2}-6p_{3/2}$	&  1.338	&  1.36(4)	&  \cite{Kastberg}	&  1.366	&  	&   \\
	&  	&  1.54(19)	&   \cite{Davidson}&  	&  	&   \\
$5d_{5/2}-6p_{3/2}$	&  4.127	&  4.15(20)	&  \cite{Kastberg}	&  4.271	&  	& \\ 
 \end{tabular}%
\end{ruledtabular}%
    \label{tab:MEvsExpBaYb}%
  \end{table}%

We present $E1$ reduced matrix elements for Ba$^+$ and Yb$^+$ in Table~\ref{tab:MEvsExpBaYb}, along with experimental values for comparison where available.
This demonstrates very good agreement between our calculations and experiment for Ba$^+$, and reasonably good agreement for Yb$^+$.
The discrepancies for the Yb$^+$ values, on the order of 5\% -- 10\%, are due mainly to the more complicated electron structure due to the closeness of the $4f^{14}$ core shell to the valence $6s$ state.
The most important $E1$ transition for the $6s$-$5d_{3/2}$ PNC amplitude in Yb$^+$ is the $p_{1/2}$-$d_{3/2}$ transition. This transition corresponds to the weak $s$-$p_{1/2}$ mixing, which dominates the amplitude. 
This $p_{1/2}$-$d_{3/2}$ $E1$ matrix element agrees with experiment to about 4\%. 
However, for the $6s$-$5d_{5/2}$ PNC amplitude considered here, the most important $E1$ amplitudes are the $s$-$p_{3/2}$ and $p_{3/2}$-$d_{5/2}$ transitions. 
The $s$-$p_{3/2}$ amplitude agrees to only 13\% with experiment, and an experimental value for the $p_{3/2}$-$d_{5/2}$ transition is, to the best of our knowledge, not known.


  \begin{table*}%
    \centering%
    \caption{Calculated magnetic dipole hyperfine constants $A$ (MHz) for the lowest valence states of Cs, Ba$^+$ and Yb$^+$, and a comparison with experiment.} 
\begin{ruledtabular}%
  \begin{tabular}{lrrlrrlrrl}
& \multicolumn{3}{c}{${}^{133}$Cs} 
& \multicolumn{3}{c}{${}^{135}$Ba$^+$} 
& \multicolumn{3}{c}{${}^{171}$Yb$^+$}  \\
\cline{2-4}\cline{5-7}\cline{8-10}
  \multicolumn{1}{c}{Level}
& \multicolumn{1}{c}{Calc.}
& \multicolumn{2}{c}{ Exp.	}
& \multicolumn{1}{c}{Calc.}
& \multicolumn{2}{c}{ Exp.	}
& \multicolumn{1}{c}{Calc.}
& \multicolumn{2}{c}{ Exp.	} \\
\hline
\smallspace
$s_{1/2}$	&  2315	&  2298.2	&  \cite{Arimondo}	&  3674	&  3593.3(22)	&  \cite{Wendt}	&  13202	&  12645(2)	&  \cite{Martensson-Pendrill}\\
$p_{1/2}$	&  290	&  291.89(8)	&  \cite{RafacTanner}	&  668	&  664.6(3)	&  \cite{Villemoes}	&  2515	&  2104.9(13)	&  \cite{Martensson-Pendrill}\\
 \end{tabular}%
\end{ruledtabular}%
    \label{tab:hfs}%
  \end{table*}%

The accuracy of the weak-charge and anapole-moment induced PNC interaction matrix elements relies on the accuracy of the wavefunctions at short distances (near the nucleus). 
One way to test the accuracy of the wavefunctions at this distance scale is to calculate magnetic dipole hyperfine structure constants, which also depend on the wavefunctions close to the nucleus. 
The hyperfine structure constants are typically reproduced very well for $s$ and $p$ states, but not so well for $d$ states (see, e.g.~Ref.~\cite{Dzuba2011}).
The direct applicability of using hyperfine structure calculations as a test for $p$-$d$ $ h_{\rm PNC}$ matrix elements has not been fully investigated, and will be the focus of future work.
The uncertainty in the calculations of the hyperfine structure constants is dominated by core polarization, which is much larger for the hyperfine constants than for the weak matrix elements.
The implication of this is that the accuracy of the $s$-$p$ PNC interaction matrix elements can be high, and importantly can be controlled by computing hyperfine constants.
For the $p$-$d$ weak matrix elements, however, there is no guarantee of high accuracy, and it is not clear how the accuracy can be reliably judged.
In Table~\ref{tab:hfs} we present calculations of magnetic-dipole hyperfine structure constants $A$, for the $6s$ and $6p_{1/2}$ states of Cs, Ba$^+$ and Yb$^+$, along with experimental values for comparison.

The $h_{\rm PNC}$ interaction, to lowest order, is effectively a contact interaction and as such only significantly mixes $s$ and $p_{1/2}$ states.
Due to core polarization, however, mixing  between $s$ and $p_{3/2}$ states, as well as between  $p_{3/2}$ and $d_{3/2,5/2}$ states, is not so small.
For $s$-$s$ PNC amplitudes there is nothing to worry about, since these contain only terms involving $s$-$p_{1/2}$ mixing.
The $s$-$d_{3/2}$ amplitudes contain also terms involving $p$-$d$ mixing, however the $s$-$p_{1/2}$ mixing is many times larger, meaning that these amplitudes are dominated by the $s$-$p_{1/2}$ mixing terms, which contribute between 70\% and 90\% to the total amplitude.

For the spin-independent amplitudes ($s$-$d_{3/2}$), the accuracy should be about 1-2\% (see Ref.~\cite{Fr-like}). 
This is due to the very good agreement with energy levels, hyperfine structure constants and matrix elements.
The spin-dependent parts of the $s$-$d_{3/2}$ amplitudes are likely to be somewhat less accurate, due mainly to core-polarization effects and the larger number of contributing states (since the spin-dependent PNC interaction can mix states with $\Delta J=1$). 
Because of this, without the double-core-polarization contribution, the accuracy for these amplitudes is likely to be between 5\% and 10\%.

For the $s$-$d_{5/2}$ amplitudes, there are no $s$-$p_{1/2}$ mixing terms, instead there are terms involving $s$-$p_{3/2}$ and $p_{3/2}$-$d_{5/2}$  mixing.
Due to core-polarization, there is no significant difference between the extent of the PNC mixing between these two contributions, and the size of the respective matrix elements is roughly the same.
For the part of these PNC amplitudes coming from the $s$-$p_{3/2}$ mixing, i.e.~the first term in Eq.~(\ref{eq:pnc}), the accuracy is likely to be good. 
However, for the contribution from the $6p_{3/2}$-$5d_{5/2}$  $\hat h_{\rm SD}$ matrix elements the accuracy is likely to be significantly worse.

There is not enough information to determine reliably how accurate the $p_{3/2}$-$d_{5/2}$ $\hat h_{\rm SD}$ matrix elements are, and as such the $s$-$d_{5/2}$ SD-PNC amplitudes should be considered order-of-magnitude estimates.
This low level of accuracy is sufficient for the purpose of the current work, which is to demonstrate the magnitude and relative sizes of these transitions in different elements. 
Note also that the very high accuracy that is required of the SI-PNC calculations for the extraction of the nuclear weak charge is not required in the search for anapole moments.

In Table~\ref{tab:compare} we compare our calculations of the SD-PNC amplitudes in Ba$^+$ and Ra$^+$ with several of those available in the literature. 
The agreement between results for the $s$-$d_{3/2}$ transitions is reasonable.
For the $s$-$d_{5/2}$ we agree with calculations of Ref.~\cite{Geetha1998} but not of Ref.~\cite{Sahoo2011}.

For atoms and ions similar to Yb$^+$, in which an external electron is close to the core and strongly interacts with its electrons, a different higher-order effect described by the so-called `ladder diagrams'~\cite{ladder} becomes important. 
The inclusion of ladder diagrams also significantly improves the accuracy of calculations in ions, for which the valence electrons lie closer to the core, and improves the accuracy of the $d$-states for atoms and ions, see e.g.~\cite{Dzuba13,Fr-like}.
With the inclusion of ladder diagrams, as well as the double-core-polarization effect, the accuracy for these calculations can potentially approach the level of several percent, though this would need further investigation.
The accuracy could then be further improved by including the Breit~\cite{Breit} and QED~\cite{QED} corrections, as well as higher order non-Brueckner electron correlations, such as structure radiation, the weak correlation potential and renormalization of states (see e.g. Ref.~\cite{CPM2,CsPNC02}).

  \begin{table}%
    \centering%
    \caption{ Reduced matrix elements $\bra{J_b,F_b}|d_{\rm SD}|\ket{J_a,F_a}$  of the spin-dependent PNC amplitudes of Ba$^+$ and Ra$^+$ and comparison with other works~\cite{NoteDefinition}.
 Units: $10^{-13}ea_0\varkappa$.} 
\begin{ruledtabular}%
  \begin{tabular}{ldcddl}
  &    &   \multicolumn{4}{c}{$E_{\rm PNC}$}  \\
\cline{3-6}
\smallspace
  & I &  \multicolumn{1}{c}{Transition}  &  \multicolumn{1}{c}{This work}  &  \multicolumn{2}{c}{Others}  \\
\hline
\smallspace
$^{135}$Ba$^+$  &  1.5  &  $\bra{5d_{5/2},3}|d_{\rm SD}|\ket{6s,2}$  &  0.85 & 0.82&  \cite{Geetha1998}\\
					&		&											&		&	0.274	&\cite{Sahoo2011}\\
\smallspace
					  & 	 &  $\bra{5d_{3/2},3}|d_{\rm SD}|\ket{6s,2}$  &  17.15 & 19.44 &  \cite{Sahoo2011}\\
\smallspace
$^{223}$Ra$^+$  &  1.5  &  $\bra{6d_{5/2},3}|d_{\rm SD}|\ket{7s,2}$  &  11.4 & 12.7& \cite{Geetha1998} \\
					&		&											&		&	3.504	&\cite{Sahoo2011}\\
\smallspace
					  & 	 &  $\bra{5d_{3/2},3}|d_{\rm SD}|\ket{6s,2}$  & 210.9  & 234.690 &  \cite{Sahoo2011}\\
  \end{tabular}%
\end{ruledtabular}%
    \label{tab:compare}%
  \end{table}%

\section{Results and discussion}\label{sec:discuss}

  \begin{table}%
    \centering%
    \caption{  SD-PNC amplitudes of the $\ket{5sF_a}\to\ket{4d_{5/2}F_b}$ transition in Rb, and the $\ket{6sF_a}\to\ket{5d_{5/2}F_b}$ transitions in Cs, Ba$^+$ and Yb$^+$. 
 Both the reduced matrix elements (RME) and the $z$ components are shown. 
Units: $10^{-13}ea_0\varkappa$. } 
\begin{ruledtabular}%
  \begin{tabular}{lllldd}
  &   &    &    &    \multicolumn{2}{c}{$E_{{\rm PNC}}$}  \\
\cline{5-6}
\smallspace
  &  $I$  &  $F_a$  &  $F_b$  &  \multicolumn{1}{r}{RME} &  \multicolumn{1}{r}{$z$-component}  \\
\hline
\smallspace
$^{85}$Rb  &  2.5  &  2  &  1  &  0.224  &  0.0708  \\
  &    &  2  &  2  &  0.409  &  0.149  \\
  &    &  2  &  3  &  0.448  &  -0.0977  \\
  &    &  3  &  2  &  0.219  &  0.0477  \\
  &    &  3  &  3  &  0.501  &  0.164  \\
  &    &  3  &  4  &  0.733  &  -0.122  \\
\smallspace
$^{87}$Rb  &  1.5  &  1  &  1  &  0.273  &  0.112  \\
  &    &  1  &  2  &  0.417  &  -0.132  \\
  &    &  2  &  1  &  0.122  &  0.0386  \\
  &    &  2  &  2  &  0.417  &  0.152  \\
  &    &  2  &  3  &  0.746  &  -0.163  \\
\smallspace
$^{133}$Cs  &  3.5  &  3  &  2  &  3.40  &  0.743  \\
  &    &  3  &  3  &  5.03  &  1.65  \\
  &    &  3  &  4  &  4.89  &  -0.815  \\
  &    &  4  &  3  &  2.91  &  0.484  \\
  &    &  4  &  4  &  5.78  &  1.72  \\
  &    &  4  &  5  &  7.71  &  -1.04  \\
\smallspace
$^{135}$Ba$^+$  &  1.5  &  1  &  1  &  -0.311  &  -0.127  \\
  &    &  1  &  2  &  -0.475  &  0.150  \\
  &    &  2  &  1  &  -0.139  &  -0.0440  \\
  &    &  2  &  2  &  -0.475  &  -0.174  \\
  &    &  2  &  3  &  -0.850  &  0.186  \\
\smallspace
$^{171}$Yb$^+$  &  0.5  &  1  &  2  &  -11.3  &  3.57  \\
\smallspace
$^{173}$Yb$^+$  &  2.5  &  2  &  1  &  -2.67  &  -0.845  \\
  &    &  2  &  2  &  -4.88  &  -1.78  \\
  &    &  2  &  3  &  -5.34  &  1.17  \\
  &    &  3  &  2  &  -2.61  &  -0.569  \\
  &    &  3  &  3  &  -5.98  &  -1.96  \\
  &    &  3  &  4  &  -8.75  &  1.46  \\
  \end{tabular}%
\end{ruledtabular}%
    \label{tab:sd5on2-Cs}%
  \end{table}%

  \begin{table}%
    \centering%
    \caption{  SD-PNC amplitudes of the $\ket{7sF_a}\to\ket{6d_{5/2}F_b}$ transitions in Fr, Ra$^+$ and Ac$^{2+}$.
 Units: $10^{-13}ea_0\varkappa$.} 
\begin{ruledtabular}%
  \begin{tabular}{lllldd}
  &   &    &    &    \multicolumn{2}{c}{$E_{{\rm PNC}}$}  \\
\cline{5-6}
\smallspace
  &  $I$  &  $F_a$  &  $F_b$  &  \multicolumn{1}{r}{RME} &  \multicolumn{1}{r}{$z$-component}  \\
\hline
\smallspace
$^{211}$Fr  &  4.5  &  4  &  3  &  24.3  &  4.05  \\
  &    &  4  &  4  &  32.6  &  9.72  \\
  &    &  4  &  5  &  29.6  &  -3.99  \\
  &    &  5  &  4  &  19.7  &  2.65  \\
  &    &  5  &  5  &  36.3  &  10.0  \\
  &    &  5  &  6  &  45.8  &  -5.18  \\
\smallspace
$^{221}$Fr  &  2.5  &  2  &  1  &  13.2  &  4.17  \\
  &    &  2  &  2  &  24.1  &  8.79  \\
  &    &  2  &  3  &  26.4  &  -5.76  \\
  &    &  3  &  2  &  12.9  &  2.81  \\
  &    &  3  &  3  &  29.5  &  9.65  \\
  &    &  3  &  4  &  43.2  &  -7.20  \\
\smallspace
$^{223}$Fr  &  1.5  &  1  &  1  &  16.1  &  6.57  \\
  &    &  1  &  2  &  24.6  &  -7.77  \\
  &    &  2  &  1  &  7.20  &  2.28  \\
  &    &  2  &  2  &  24.6  &  8.97  \\
  &    &  2  &  3  &  44.0  &  -9.59  \\
\smallspace
$^{223}$Ra$^+$  &  1.5  &  1  &  1  &  4.16  &  1.70  \\
  &    &  1  &  2  &  6.35  &  -2.01  \\
  &    &  2  &  1  &  1.86  &  0.588  \\
  &    &  2  &  2  &  6.35  &  2.32  \\
  &    &  2  &  3  &  11.4  &  -2.48  \\
\smallspace
$^{225}$Ra$^+$  &  0.5  &  1  &  2  &  14.4  &  -4.55  \\
\smallspace
$^{229}$Ra$^+$  &  2.5  &  2  &  1  &  3.41  &  1.08  \\
  &    &  2  &  2  &  6.22  &  2.27  \\
  &    &  2  &  3  &  6.82  &  -1.49  \\
  &    &  3  &  2  &  3.33  &  0.726  \\
  &    &  3  &  3  &  7.62  &  2.49  \\
  &    &  3  &  4  &  11.2  &  -1.86  \\
\smallspace
$^{227}$Ac$^{2+}$  &  1.5  &  1  &  1  &  4.59  &  1.88  \\
  &    &  1  &  2  &  7.02  &  -2.22  \\
  &    &  2  &  1  &  2.05  &  0.650  \\
  &    &  2  &  2  &  7.02  &  2.56  \\
  &    &  2  &  3  &  12.6  &  -2.74  \\
  \end{tabular}%
\end{ruledtabular}%
    \label{tab:sd5on2-Fr}%
  \end{table}%

Our calculations of the $s$-$d_{5/2}$ SD-PNC amplitudes of several isotopes of Rb, Cs, Ba$^+$ and  Yb$^+$ are presented in Table~\ref{tab:sd5on2-Cs}, and for Fr, Ra$^+$ and Ac$^{2+}$ in Table~\ref{tab:sd5on2-Fr}.
For ease of comparison we present both the reduced matrix elements, 
defined in Eq.~(\ref{eq:sd-pnc-rme}), and the $z$-components. 
The $s$-$d_{5/2}$ are typically between one and two orders of magnitude smaller than the corresponding $s$-$d_{3/2}$ transitions, due primarily to the absence of $s$-$p_{1/2}$ weak mixing.
The largest amplitudes presented are in Fr, consistent with its very large $s$-$s$ and $s$-$d_{3/2}$ transitions.
The amplitudes are large in fact for all the Fr-like ions, and are also large in Cs and Yb$^+$.

As well as the $s$-$d_{5/2}$  transitions, which have no SI contribution, we have also performed calculations for several $s$-$d_{3/2}$ transitions for which both SI and SD contributions are non-zero.
We express these amplitudes in the form
$E_{\rm PNC} = P(1+R),$
where $P$ is the SI PNC amplitude (including $Q_W$), and $R$ is the ratio of the SD to SI parts.
Here we calculate both parts concurrently, using the same method and wavefunctions. 
This approach has the advantage that the relative sign difference between the SI and SD parts is fixed, ensuring no ambiguity in the sign of  $\varkappa$~\cite{Dzuba2011Yb}.
There is also typically a significant improvement in accuracy for the ratio over that for each of the amplitudes individually, due to the fact that the atomic calculations for both components are very similar and much of the theoretical uncertainty cancels in the ratio~\cite{Dzuba2011}.

We present these amplitudes for Rb and Cs in Table~\ref{tab:sd3on2-Cs}, and for Fr and Ac$^{2+}$ in Table~\ref{tab:sd3on2-Fr}.
We don't present amplitudes for Ba$^{+}$, Yb$^{+}$ or Ra$^{+}$ since these have been performed in our recent work Ref.~\cite{Dzuba2011}.

  \begin{table}%
    \centering%
    \caption{
	PNC amplitudes ($z$~components) of  the $\ket{5sF_a}\to\ket{4d_{3/2}F_b}$ transition in Rb, and the $\ket{6sF_a}\to\ket{5d_{3/2}F_b}$ transitions in Cs.
	 Units: $10^{-11}ea_0$.
} 
\begin{ruledtabular}%
  \begin{tabular}{ldlllr}
  & Q_W & $I$  &  $F_a$  &  $F_b$  &   \multicolumn{1}{c}{$E_{\rm PNC}$}  \\
\hline
\smallspace
$^{87}$Rb  &  -46.8  &  1.5  &  1  &  0  &  $ -0.301 	\times[1+0.0805\varkappa]$ \\
  &    &    &  1  &  1  & $ -0.337 	\times[1+0.0796\varkappa]$ \\
  &    &    &  1  &  2  & $ 0.261 	\times[1+0.0779\varkappa]$ \\
  &    &    &  2  &  1  & $ -0.117 	\times[1-0.0439\varkappa]$ \\
  &    &    &  2  &  2  & $ -0.301 	\times[1-0.0457\varkappa]$ \\
  &    &    &  2  &  3  & $ 0.301	 \times[1-0.0483\varkappa]$ \\
 \smallspace
$^{133}$Cs  &  -73.2  &  3.5  &  3  &  2  &  $-2.05 	\times[1+0.0444\varkappa] $\\
  &    &    &  3  &  3  &  $-3.14 	\times[1+0.0431\varkappa]$ \\
  &    &    &  3  &  4  & $ 1.35 		\times[1+0.0412\varkappa] $\\
  &    &    &  4  &  3  & $ -0.923 	\times[1-0.0305\varkappa]$ \\
  &    &    &  4  &  4  & $ -2.86 	\times[1-0.0323\varkappa] $\\
  &    &    &  4  &  5  & $ 1.87	 	\times[1-0.0345\varkappa] $\\
  \end{tabular}%
\end{ruledtabular}%
    \label{tab:sd3on2-Cs}%
  \end{table}%

  \begin{table}%
    \centering%
    \caption{
	PNC amplitudes of the $\ket{7sF_a}\to\ket{6d_{3/2}F_b}$ transitions in Fr  and Ac$^{2+}$.
	 Units: $10^{-11}ea_0$.
} 
\begin{ruledtabular}%
  \begin{tabular}{ldlllr}
  & Q_W & $I$  &  $F_a$  &  $F_b$  &   \multicolumn{1}{c}{$E_{\rm PNC}$}  \\
\hline
\smallspace
$^{223}$Fr  &  -128.3  &  1.5  &  1  &  0  & $ -38.4 	\times[1+0.0273\varkappa]$ \\
  &    &    &  1  &  1  &$  -43.0 	\times[1+0.0278\varkappa] $\\
  &    &    &  1  &  2  &$  33.3 		\times[1+0.0288\varkappa]$ \\
  &    &    &  2  &  1  &$  -14.9 	\times[1-0.0189\varkappa]$ \\
  &    &    &  2  &  2  &$  -38.4 	\times[1-0.0179\varkappa]$ \\
  &    &    &  2  &  3  &$  38.4 		\times[1-0.0164\varkappa] $\\
  \smallspace
$^{227}$Ac$^{2+}$  &  -130.1  &  1.5  &  1  &  0  & $ -28.7	\times[1+0.0250\varkappa] $\\
  &    &    &  1  &  1  &$  -32.0 	\times[1+0.0241\varkappa]$ \\
  &    &    &  1  &  2  &$  24.8 		\times[1+0.0223\varkappa]$ \\
  &    &    &  2  &  1  &$  -11.1 	\times[1-0.0105\varkappa]$ \\
  &    &    &  2  &  2  &$  -28.7 	\times[1-0.0123\varkappa]$ \\
  &    &    &  2  &  3  &$  28.7 		\times[1-0.0150\varkappa] $\\
  \end{tabular}%
\end{ruledtabular}%
    \label{tab:sd3on2-Fr}%
  \end{table}%

\subsection{Suitability for measurements}

A method has been proposed by Fortson for measuring PNC in a single atomic ion that has been laser trapped and cooled~\cite{Fortson}.
Originally proposed with measuring the $6s$-$5d_{3/2}$ transition of Ba$^+$ in mind, work has begun to use this method for the $7s$-$6d_{3/2}$ transition in  Ra$^+$ at KVI~\cite{KVI}.
The use of this or a similar method to study spin-dependent PNC in $s$-$d_{5/2}$ transitions has been previously discussed~\cite{Geetha1998,Sahoo2011,propose}.
Though these transitions have significantly smaller PNC signals than the corresponding $s$-$d_{3/2}$ transitions, the main advantage here is that there is no SI contribution. 
This is beneficial for the extraction of the nuclear anapole moment since the (larger) SI contribution would not need to be subtracted, and it would limit the possibility of spurious SI-PNC acting as a false signal.

In~\cite{Fortson} it was shown that to ensure accurate PNC measurements of a single trapped ion both the upper and lower levels of the transition should be long lived.
The only significant contribution to the decay rate of the $5,6d_{5/2}$ states in Ba$^{+}$, Ra$^{+}$ are the E2 transitions to the $s$ ground state. 
There are also M1 and E2 $d_{5/2}$-$d_{3/2}$  contributions, though these are highly suppressed.
Both E2 transitions are suppressed in the case of Ac$^{2+}$, so we include both in the calculation.
We calculate the lifetimes of the relevant $d_{5/2}$ states in Ba$^{+}$, Ra$^{+}$ and Ac$^{2+}$ to be 35.9 s, 0.302 s, and 247 s respectively.
These results are in good agreement with other recent calculations, e.g.~\cite{Pal2009,Sahoo2006}.
The upper states of the other elements presented here are unstable as they have allowed E1 transitions to lower levels. This is not a problem for neutral Cs or Fr where atomic-beam-type experiments could be used.

In the $s$-$d_{5/2}$ transitions considered here it is possible  that the contribution to the amplitude coming from the combination of the weak charge and hyperfine interaction may not be as small as in other systems, due to the $d_{3/2}$-$d_{5/2}$ and $p_{1/2}$-$p_{3/2}$ hyperfine mixing.
The ratio of the hyperfine to fine structure splitting goes as 
\begin{equation}
\frac{1}{Z}\frac{m_e}{m_p}\sim10^{-5}.
\end{equation}
The PNC amplitude of the $s$-$d_{5/2}$ transitions due to the combined weak charge and hyperfine interaction would therefore be of the order
\begin{equation}
E_{\rm PNC}^{Q_W+{\rm hf}}(s{\rm-}d_{5/2}) \sim 10^{-5}\, E_{\rm PNC}^{Q_W}(s{\rm-}d_{3/2}).
\end{equation}
For Cs, this leads to a $Q_W+{\rm hf}$ contribution on the order of $10^{-16}$ (including $Q_W$), whereas the anapole moment contribution to this transition is $10^{-13}\varkappa\sim 10^{-14}$.
Similarly for Ba$^+$, Fr and Ra$^+$ the $Q_W+{\rm hf}$ contribution is between one and two orders of magnitude smaller than the contribution from the anapole moment. This is smaller than the assumed accuracy here, so this contribution can be safely neglected for now.
An accurate calculation of this  contribution is beyond the scope of the current work, and will be the focus of a future study.

\section{Conclusion}

We have presented order-of-magnitude calculations of nuclear-spin-dependent PNC amplitudes for the $s$-$d_{5/2}$ transitions of several heavy atoms and ions.
Also presented are PNC amplitudes of the $s$-$d_{3/2}$ transitions of the same ions (where not presented previously) that are accurate to the $\sim10\%$ level.
These calculations could be used to extract an experimental value of the nuclear anapole moment, which in turn could be used to study parity violating nuclear forces.
The accuracy of these calculations could be improved with the inclusion of higher order correlation corrections, such as the double-core-polarization~\cite{DCP}, structure radiation~\cite{CPM2} and ladder-diagrams~\cite{ladder}, as well as other small corrections such as the Breit~\cite{Breit} and QED~\cite{QED} corrections.

\acknowledgments
One of the  authors (V.A.D.) would like to express a special thanks to
the Mainz Institute for Theoretical Physics (MITP) for its hospitality
and support. 
We extend our thanks to D. S. Elliot for stimulating this work.
The work was also supported by the Australian Research Council.



\begin{thebibliography}{40}
\bibitem{NoteDefinition} 
Note that several different definitions of $\varkappa$ and the SD-PNC Hamiltonian exist in the literature. The reader should be aware that these differences must be taken into account when comparing results.
\end{thebibliography}
\end{document}